\documentclass[pra,aps,twocolumn]{revtex4-1}
\usepackage{graphicx}
\usepackage{float}
\usepackage{amssymb}

\begin{document} 
\title{Effect of defects on the phonons and the effective spin-spin interactions of an ultracold Penning
trap quantum simulator}

\author{M. McAneny}
\affiliation{Department of Physics, Georgetown University, 37th and O Sts. NW,
Washington, DC 20057, USA}

\author{B. Yoshimura}
\affiliation{Department of Physics, Georgetown University, 37th and O Sts. NW,
Washington, DC 20057, USA}

\author{J. K. Freericks}
\affiliation{Department of Physics, Georgetown University, 37th and O Sts. NW,
Washington, DC 20057, USA}

\date{\today}


\def\bx{{\bf x}}
\def\bk{{\bf k}}
\def\br{{\bf r}}
\def\bu{{\bf u}}
\def\half{\frac{1}{2}}
\def\args{(\bx,t)}

\begin{abstract}

We generalize the analysis of the normal modes for a rotating ionic
Coulomb
crystal in a Penning trap to allow for inhomogenieites in the system. Our formal developments
are completely general, but we choose to examine a crystal of Be$^+$ ions with BeH$^+$ defects to
compare with current experimental efforts. We examine the classical phonon modes (both transverse and planar) and we determine the effective spin-spin interactions when the system is driven by an axial spin-dependent optical dipole force.
We examine situations with up to approximately 15\% defects.
We find that most properties are not strongly influenced by the defects, indicating that the presence of a small number of defects will not significantly affect experiments.

\end{abstract}
\maketitle

\section{Introduction}

Current work on quantum-mechanical simulators is 
motivated by Feynman's 1982 proposal~\cite{feynman}.
One of the most successful systems for engineering
model spin systems (which evolve naturally and whose properties can be measured
directly) are ion trap emulators~\cite{two-ion,Kim1,Kim2,Edwards,Islam,John,John2,Blatt,Islam2}. These systems
are advantageous because of their long decoherence times, precise spin-state
quantum control and high-fidelity readout. But most experiments have not yet reached system sizes
where they are superior to conventional digital computer
calculations (for example, linear Paul trap experiments have worked with as many as 16 spins~\cite{Islam2} in a one-dimensional linear crystal).
However, using a Penning trap, researchers at NIST in Boulder, CO have created systems
of
$\sim$300 spins in a single-plane two-dimensional crystal and have demonstrated the use of a spin-dependent optical dipole
force to realize tunable Ising-type spin-spin couplings, an important first step in the
simulation of quantum magnetism in systems intractable by conventional computers~\cite{John,John2}.

One complication to the NIST experiment is the accumulation of impurities
in
the crystal during the experimental procedure. Hydrogen molecules and water molecules in the
background
gas collide with the Beryllium ions in the crystal and form a variety
of impurities~\cite{wineland}, the most common being BeH$^+$ ions or BeOH$^+$ ions (H$^+$ ions can also form as defects if the electron impact ionization electrons have too high an energy or the system is at too high a temperature, but when that happens the crystal is usually reloaded, so we neglect them). In this article, we
generalize
previous work for the normal modes in a Penning trap~\cite{georgetown} to include these types of
defects. In our numerical calculations, we focus solely on the BeH$^+$ defects (which are the most commonly seen defect), although the generalization to other defects is a straightforward and simple task. Because these defects are more massive than the Be$^+$ ions, they are centrifugally driven to the boundary of the planar array.
We determine how these defects affect the eigenfrequencies and eigenvectors of the
normal modes,
and investigate how these changes affect  the spin-spin interactions when
detuning the optical dipole force to the blue of the axial center-of-mass mode.

The organization of the paper is as follows.
In Sec. II, we generalize the theory for the normal modes of cold ions in a
Penning
trap to include the  presence of impurities. 
In Sec. III, we provide representative numerical examples to illustrate how BeH$^+$
defects modify the phonon frequencies and the spin-spin interactions.
In Sec. IV, we provide our conclusions.

\begin{figure}[http!]
\centering
\includegraphics[scale=0.4]{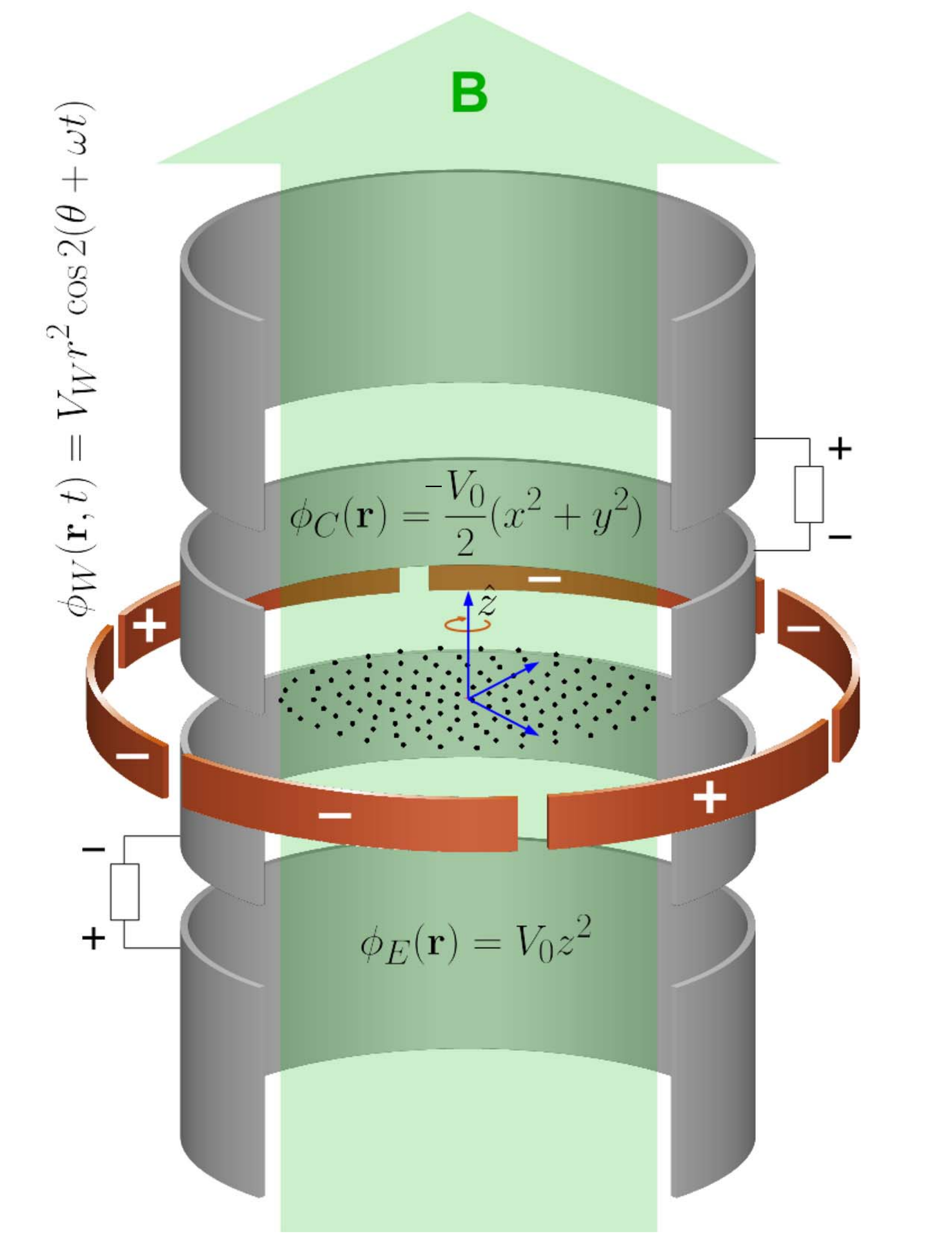}\\
\caption{
Electrostatic potentials  provided by electrodes in a Penning trap have
contributions from both the end-cap electrodes [with the trapping potential
$\phi_{E}({\bf r})=V_{0}z^{2}$ that pushes the atoms towards $z=0$] and the
cylindrical electrodes [with the radial quadratic potential $\phi_{C}({\bf
r})=-\frac{V_{0}}{2}(x^2+y^2)$, which tends to push the ions out radially]. In
addition, the repulsive Coulomb potential between the ions tends to destabilize
the system in the trap.
A static magnetic field ${\bf B}=B_{z}{\hat z}$($B_{z}>0$) is thus applied along
the axial direction to provide the radial confinement of the ions.
To lock the rotational angular frequency of the ions at a specific rotational
speed, a time-dependent clockwise quadrupole potential $\phi_{W}(t)=V_{W} r^{2}
\cos2(\theta +\omega t)$ ($\omega > 0$) is applied to the ions through ring
electrodes so that the steady state of the ions (with a rigid body rotational
speed $\omega)$ can be phase locked. Note that the rotating quadrupole potential
is well implemented by ring electrodes with just six sectors (as employed in the
NIST Penning
traps~\cite{John,John3,John4}).}
  \label{fig:FIG1}
\end{figure}

\section{Theoretical Formulation for a Penning trap with defects}
We consider a collection of ions formed as a two-dimensional
crystal in a Penning trap as shown in Fig.~1. Previous work~\cite{georgetown}
considered a homogeneous crystal of $^{9}\textrm{Be}^{+}$ ions, constructing
the Lagrangian for the system and transforming to a co-rotating frame to remove the 
time dependence from the Lagrangian in order to treat the problem as a static
equilibrum problem.
When allowing for the possibility of defects, the only change to this
formulation is to include an index for the mass degree of freedom, allowing it to change from site to site. Thus, the Lagrangian in the rotating frame for any collection of
ions is (the superscript $R$ denotes the rotating frame)
\begin{equation}
L^{R}=\sum_{j=1}^{N}\left[\frac{1}{2} m_j|\dot{{\bf
r}}_{j}^{R}|^{2}-\frac{eB^{eff}[j,\omega]}{2}({\dot x}_{j}^{R}y_{j}^{R}-{\dot
y}_{j}^{R}x_{j}^{R})-e\phi_{j}^{R}\right]
\label{eq:Lagrangian}
\end{equation}
where $N$ is the total number of ions, $B^{eff}[j,\omega]=B_{z}-2\omega m_j/e$ is the effective magnetic field in the rotating frame,
 $\omega>0$ is the angular rotation frequency of the crystal ($\Omega = -\omega \hat{z}$),
$m_j$ is the mass of ion $j$, $e$ is the positive unit charge, and  ${\bf
r}_{j}^{R}$ is the position vector from the origin of the plane to the $j$th ion.  We use both Cartesian coordinates ${\bf r}_j^R=(x_j^R,y_j^R,z_j^R)$ and cylindrical coordinates, where $\rho_j^R=\sqrt{(x_j^{R})^2+(y_j^{R})^2}$, in our formulas. The effective potential satisfies
\begin{eqnarray}
e\phi_{j}^{R} &=&
eV_{0}{z_{j}^{R}}^{2}+\frac{1}{2}(eB_{z}\omega-m_j\omega^{2}-eV_{0}){\rho_{j}^{R}}^
{2 } \nonumber \\
 && + eV_{W}({x_{j}^{R}}^{2}-{y_{j}^{R}}^{2}) +\frac{k_ee^{2}}{2}\sum_{k_e\neq
j}\frac{1}{ r_{jk}^{R}}
\end{eqnarray} 
with $V_0$ the amplitude of the Penning trap electrodes, $V_W$ the amplitude of
the rotating wall potential,
and $r_{jk}^{R}=|{\bf r}_j^R-{\bf r}_k^R|$ the interparticle distance between ions $j$ and $k$.
In this frame, the centrifugal force originates from the term
$-\frac{1}{2}m\omega^2{\rho_{j}^R}^2$
in the effective potential energy $e\phi_{j}^{R}$.
 Details for how to derive this formula can be found from a simple generalization of the derivation in Ref.~\cite{georgetown}.

\subsection{Equilibrium Structure and Normal Modes in the Rotating Frame}
We solve for the equilibrium structure of the crystal in a standard way~\cite{georgetown,chinese,Jake},
constructing initial ``shell" configurations and minimizing the effective potential
in the rotating frame. We express all distances in terms of a characteristic length, denoted $\ell_0$, which satisfies
\begin{equation}
\ell_0=\sqrt[3]{\frac{2ke^2}{m_{Be^+}\omega_z^2}}=\sqrt[3]{\frac{ke^2}{eV_0}}.
\end{equation}
However, for crystals with defects, we solve the problem in a three-step fashion: (i) first we solve the pure case, with all masses equal, and then (ii) we
replace the ion located farthest from the center with a defect, re-solve for the stable equilibrium positions starting from the current ion positions, and then (iii) iteratively continue
until the chosen number of total defects, $N_d$ is reached. We choose this procedure when the defect ions are heavier than the majority ions, since the centrifugal force will push the heavier ions to the periphery of the crystal.  (A simple generalization, which places lighter ions in positions closest to the center of the trap would be used for lighter defects like H$^+$ ions if one wanted to study such a system.)

After solving for the equilibrium structure given by the equilibrium lattice site positions $\{{\bf R}_j^0, j=1\ldots,n\}$, we expand the Lagrangian about this configuration to quadratic order with ${\bf r}_j^R={\bf R}_j^0+\delta {\bf R}_j(t)$ (see Ref.~\cite{georgetown} for details).
The Lagrangian governing the axial motion decouples
from the Lagrangian of the planar motion, yielding
\begin{equation}
L_{ph}^{A}= \frac{1}{2}\sum_{j=1}^{N}m_j{\delta\dot{R}_{j}^{z}}{\delta\dot{R}_{j}^{z}}
-\frac{1}{2}\sum_{j,k=1}^{N}{K}_{jk}^{zz}\delta R_{j}^{z}\delta R_{k}^{z},
\end{equation}
in which the symmetric stiffness matrix ${\bf K}^{zz}$ is a real Hermitian matrix
 $K_{jk}^{zz}=K_{kj}^{zz}$ whose matrix elements are given by
\begin{equation}
K_{jk}^{zz}=\left\{
\begin{array}{lll}
    \displaystyle {2eV_{0}-\sum_{l=1, l\ne j}^{N}\frac{k_{e}e^{2}}{({R_{l j}^{0}})^{3}}} &\ \ \ \ j=k,\\
    \displaystyle {\frac{k_ee^2}{({R_{jk}^{0}})^{3}}} &\ \ \ \ j \neq k,
\end{array}\right.
\end{equation}
where $R_{jk}^{0}=|{\bf R}_j^0-{\bf R}_k^0|$ is the distance between two ions in equilibrium (in the rotating frame).
Similarly, we derive the Lagrangian for the planar normal modes to be
\begin{equation}
\begin{array}{ll}
\displaystyle{L_{ph}^{P} = \frac{1}{2}\sum_{j=1}^{N}\sum_\alpha m_j(\dot{\delta R_j^\alpha})^{2}-\frac{1}{2}
\sum_{j,k=1}^{N}\sum_{\alpha\beta} K_{jk}^{\alpha\beta} \delta R_{j}^{\alpha}
\delta R_{k}^{\beta}} \\
\displaystyle{+\sum_{j}^{N}\frac{eB^{eff}[j,\omega]}{2}
\left[\delta R_{j}^{x}
\delta{\dot R}_{j}^{y}-\delta R_{j}^{y}\delta{\dot R}_{j}^{x}\right]},
\end{array}
\end{equation}
where $\alpha$ and $\beta$ run over the $x$ and $y$ degrees of freedom only.
The real Hermitian stiffness matrix ${\bf K}^{\alpha\beta}$ 
satisfies the relation $K_{jk}^{\alpha\beta}=K_{kj}^{\alpha\beta}=K_{jk}^{\beta\alpha}=K_{kj}^{\beta\alpha}$ (note that the Roman indices denote lattice sites, while the Greek indices denote the coordinate directions $x$ and $y$).
We derive the stiffness matrix ${\bf K}^{\alpha\beta}$  for the planar modes to satisfy
\[
K_{jk}^{xx}=\left\{
\begin{array}{llll}
 \displaystyle{m_j\omega^2+e\omega B^{eff}[j,\omega]-eV_{0}+2eV_{\omega}}
 &\\
  \displaystyle{-k_{e}e^{2}\sum_{l=1, l\ne j}^{N}
   \frac{({R_{jl}^{0}})^{2}-3(x_{j}^{0}-x_{l}^{0})^{2}}{
    ({R_{jl}^{0}})^{5}}} & \ \ \ \ j=k, \\
       &   \\
   \displaystyle{k_{e}e^{2}\frac{({R_{jk}^{0}})^{2}-3(x_{j}^{0}-x_{k}^{0})^{2}}
  {({R_{jk}^{0}})^{5}}} & \ \ \ \ j \neq k,
\end{array}\right.
\]
\[
K_{jk}^{yy}=\left\{
\begin{array}{llll}
 \displaystyle{m_j\omega^2+e\omega B^{eff}[j,\omega]-eV_{0}-2eV_{\omega}}
 &\\
  \displaystyle{-k_{e}e^{2}\sum_{l=1, l\ne j}^{N}
   \frac{({R_{jl}^{0}})^{2}-3(y_{j}^{0}-y_{l}^{0})^{2}}{
    ({R_{jl}^{0}})^{5}}} & \ \ \ \ j=k,\\
       &   \\
   \displaystyle{k_{e}e^{2}\frac{({R_{jk}^{0}})^{2}-3(y_{j}^{0}-y_{k}^{0})^{2}}
  {({R_{jk}^{0}})^{5}}} & \ \ \ \ j \neq k,
\end{array}\right.
\]

\begin{equation}
K_{jk}^{xy}=\left\{
\begin{array}{ll}
  \displaystyle{3k_{e}e^2\sum_{l=1, l\ne j}^{N}
  \frac{(x_{j}^{0}-x_{l}^{0})(y_{j}^{0}-y_{l}^{0})}{({R_{jl}^{0}})^{5}}} \ \ \  &  \ \ \ \ j=k,\\
  & \\
\displaystyle{-3k_{e}e^2\frac{(x_{j}^{0}-x_{k}^{0})(y_{j}^{0}-y_{k}^{0})}
{({R_{jk}^{0}})^{5}}}& \ \ \ \ j \neq k
\end{array}\right.
\end{equation}
where the equilibrium configuration is represented by the ion coordinates ${\bf R}_{j}^{0}=(x_{j}^{0}, y_{j}^{0}, z_{j}^{0})$, for $ j=1,2,...,N$.
Note that the off-diagonal matrix elements ${\bf K}^{xy}$ have nonzero values,
indicating the collective nature of the planar motional degrees of freedom, which couple the motion in the two coordinate directions together.

\subsection{Axial phonon modes}
After, applying the Euler-Lagrange equation to $L_{ph}^{A}$ and substituting the eigenvector solution 
$\delta R_j^z(t)= b_{j}^{z\nu}\cos [\omega_{z\nu}(t-t_0)]$, we are left to solve the generalized eigenvalue equation,
\begin{equation}
\sum_{k=1}^{N}\left[\omega_{z\nu}^{2}m_j
\delta_{jk}-K_{jk}^{zz}\right]b_{k}^{z\nu}=0,\ \nu=1,2, \ldots N.
\end{equation}
It is a simple procedure to convert this into a standard eigenvalue problem.  First, we define the average mass to satisfy $m_{\rm ave} =\sum_{i=1}^Nm_j/N$. Then we define the mass matrix to be $M_{jk}=m_j\delta_{jk}$ which is a diagonal matrix.  Finally, we define a new stiffness matrix via
\begin{equation}
\bar {\bf K}^{zz}=m_{\rm ave}{\bf M}^{-\frac{1}{2}}{\bf K}^{zz}{\bf M}^{-\frac{1}{2}},
\label{eq: kzz_bar_def}
\end{equation}
which has the same physical units as the original stiffness matrix and remains real and symmetric.  The inverse matrix square root operation on the mass matrix {\bf M} is trivial to compute because the mass matrix is a diagonal matrix and all of its diagonal elements are positive, so ${ M}^{-1/2}_{jk}=\delta_{jk}/\sqrt{m_j}$.
We also define a new set of eigenvectors $\bar b_j^{z\nu}=b_j^{z\nu}\sqrt{m_j}/\sqrt{m_{\rm ave}}$ so that $\bar b^{z\nu}={\bf M}^{1/2}b^{z\nu}/\sqrt{m_{\rm ave}}$. These definitions now allow us to find a conventional eigenvalue problem to solve, namely
\begin{equation}
\sum_{k=1}^{N}\left[m_{ave}\omega_{z\nu}^{2}
\delta_{jk}-\bar K_{jk}^{zz}\right]\bar b_{k}^{z\nu}=0,\ \nu=1,2, \ldots N.
\end{equation}
Since the new stiffness matrix is real and symmetric, we can choose the eigenvectors to be real. Denoting the eigenvalues of $\bar {\bf K}^{zz}$ by $\lambda_{z\nu}$, the eigenfrequencies for the axial modes become $\omega_{z\nu}=\sqrt{\lambda_{z\nu}/m_{\rm ave}}$.  The set of eigenvectors $\{\bar b^{z\nu}, \nu=1\ldots N\}$ forms an orthonormal and complete set of eigenvectors, because $\bar{\bf K}^{zz}$ is a real and symmetric matrix.  The original eigenvectors $\{b^{z\nu}=\sqrt{m_{\rm ave}}{\bf M}^{-1/2}\bar b^{z\nu}, \nu=1\ldots N\}$ will not, in general,  be orthogonal, but instead they satisfy the following orthonormality conditions:
\begin{equation}
\sum_{j=1}^N \frac{m_j}{m_{\rm ave}}b_j^{z\nu}b_j^{z\nu^\prime}=\delta_{\nu\nu^\prime},
\end{equation}
and the following completeness relation
\begin{equation}
\sum_{\nu=1}^N b_j^{z\nu}b_k^{z\nu}=\frac{m_{\rm ave}}{m_j}\delta_{jk}.
\end{equation}

In Sec. III, we will show numerical results for the eigenvalue problem, but here we want to mention one important fact.  When we have a pure system, all of the masses are identical, and one can immediately show that the center-of-mass mode involves an identical axial displacement of all of the ions, and hence it has a frequency equal to $\omega_z$. As defects are added into the system, the center-of-mass
mode changes, with the heavier mass objects moving with a smaller amplitude. This then reduces the frequency of the mode, which is one of the simplest ways to determine that the crystal has defects (in fact, by tracking the discrete changes of this mode frequency, one can determine the precise number of defects at any given instant of time). See below in Fig.~\ref{fig: com_defect} where we show how the center-of-mass frequency tracks with the number of defects.  Because the center-of-mass mode is no longer uniform, it will not generate a uniform spin-spin coupling between the different ions, which is possible only when the system has no defects.

The quantization condition is standard.  We let $\delta R_j^z(t)=\sum_\nu \xi_\nu(t) b_j^{z\nu}=\sum_\nu \xi_\nu(t)\bar b_j^{z\nu} \sqrt{m_{\rm ave}/m_j}$ be the displacements from the equilibrium positions, with the quantum dynamics due to the $\nu$th phonon mode arising from the real generalized coordinate $\xi_\nu(t)$.  The Lagrangian for the axial motion then becomes
\begin{equation}
L_{ph}^A=\frac{1}{2}\sum_{j=1}^Nm_{\rm ave}\left [ \dot\xi_\nu^2-\omega_{z\nu}^2\xi_\nu^2\right ].
\end{equation}
The generalized momentum is $P_\nu^A=m_{\rm ave}\dot \xi_\nu$, and the Hamiltonian takes the standard form, which can be expressed as
\begin{equation}
H_{ph}^A=\sum_{\nu=1}^N \hbar\omega_{z\nu}\left ( \hat n_{z\nu}+\frac{1}{2}\right ),
\end{equation}
where $\hat n_{z\nu}=a_{z\nu}^\dagger a_{z\nu}$ is the phonon number operator, expressed in terms of
the phonon raising and lowering operators, which satisfy
\begin{eqnarray}
a_{z\nu}^\dagger &=&\sqrt{\frac{m_{\rm ave}\omega_{z\nu}}{2\hbar}}\left (\xi_{z\nu}+\frac{i}{m_{\rm ave}\omega_{z\nu}}P_\nu^A\right ),\\
a_{z\nu}&=&\sqrt{\frac{m_{\rm ave}\omega_{z\nu}}{2\hbar}}\left (\xi_{z\nu}-\frac{i}{m_{\rm ave}\omega_{z\nu}}P_\nu^A\right ).
\end{eqnarray}
The raising and lowering operators have the conventional commutation relations $[a_{z\nu},a_{z\nu^\prime}^\dagger]_-=\delta_{\nu\nu^\prime}$. The axial displacement is then 
represented as an operator in the following two equivalent forms:
\begin{eqnarray}
\delta \hat R_j^z&=&\sum_\nu \sqrt{\frac{\hbar}{2m_{\rm ave}\omega_{z\nu}}}(a_{z\nu}^\dagger+a_{z\nu})b_j^{z\nu};\nonumber\cr
&=&\sum_\nu \sqrt{\frac{\hbar}{2m_j\omega_{z\nu}}}(a_{z\nu}^\dagger+a_{z\nu})\bar b_j^{z\nu}.
\end{eqnarray}

Finally, we need to derive the effective spin-spin interactions that arise when an optical dipole force is applied to the system in the axial direction.  The analysis is identical to that given before~\cite{georgetown,spin-spin-interactions}, with the mass replaced by the average mass, due to the form of the expansion of the coordinate in terms of the phonon raising and lowering operators and the $b^{z\nu}$ generalized eigenvectors. Hence, we find that the Ising spin-spin coupling between sites $j$ and $j'$ is
\begin{eqnarray}
  J_{jj'}(t) &=& \frac{F_{O}^2}{4m_{\rm ave}}\sum_{\nu=1}^{N}\frac{b_{j}^{z \nu}b_{j'}^{z \nu}}{\mu^2-\omega_{z \nu}^2} \nonumber \\
    & \times & [1+\cos{2\mu t}-\frac{2\mu}{\omega_{z\nu}}\sin{\omega_{z \nu} t}\sin{\mu t}],
  \label{eq:axialspin-spin}
\end{eqnarray}
where $F_O$ is the magnitude of the optical dipole force, and $\mu$ is the beatnote frequency corresponding to the frequency difference of the two off-resonant laser beams being applied to the trapped ion crystal. The symbol $\mu$ is also referred to as the detuning of the optical dipole force from the phonon modes, since the crystal is resonantly driven when the beatnote frequency (or detuning) is equal to a normal mode frequency $\omega_{z\nu}$. Of course, the spin-spin interactions only hold between the pure atomic species, since the energy levels and matrix elements change for the defect sites, so the sites $j$ and $j'$ must both be sites without defects, in order for there to be a spin-spin interaction generated by the optical dipole force.

\subsection{Planar phonon modes}
Now we discuss the planar phonon modes, which are more complicated.
Applying the Euler-Lagrange equations to the classical  planar Lagrangian $L_{ph}^{P}$, results in the following coupled equations of motion
in the $xy$ plane:
\begin{equation}
\label{eq:LONGITUDINAL}
\begin{array}{l}
\displaystyle{
  m_j {\delta \ddot R}_{j}^{x}+\sum_{k=1}^{N}\Big[K_{jk}^{xx}\delta R_{k}^{x}+K_{jk}^{xy}\delta R_{k}^{y}   -eB_{eff}[j,\omega]\delta_{jk}\delta\dot{R}_{k}^{y}\Big]}\\{=0;}\\
  \displaystyle{m_j\delta\ddot{R}_{j}^{y}+\sum_{k=1}^{N}\Big[K_{jk}^{yy}\delta R_{k}^{y}+K_{jk}^{yx}
  \delta R_{k}^{x}   +eB_{eff}[j,\omega]\delta_{jk}\delta \dot{R}_{j}^{x}\Big]}\\{=0 }.
\end{array}
\end{equation}
where the ion index $j$ runs from  $1$ to $N$. We assume we have a set of $2N$ planar eigenfrequencies $\omega_\lambda$ and we express the displacements at each site in terms of ``composite'' eigenvectors $\sum_{\nu=1}^{2N} \alpha_\lambda^\nu b_j^{\alpha\nu}$ where $\alpha_\lambda^\nu$ are complex eigenvectors, while $b_j^{\alpha\nu}$ are real eigenvectors, described below. The superscript $\alpha$ denotes the $x$ or the $y$ spatial components.  Hence, we let 
\begin{eqnarray}
\delta R_j^\alpha(t)&=&\sum_{\lambda=1}^{2N}\sum_{\nu=1}^{2N}\frac{1}{2} \left [ \alpha_\lambda^{\nu *} e^{i\omega_\lambda t}+\alpha_\lambda^\nu e^{-i\omega_\lambda t}\right ] b_j^{\alpha\nu},\\
&=&\sum_{\lambda=1}^{4N}\sum_{\nu=1}^{2N} \frac{1}{2}{\rm Re}\left [\alpha_\lambda^\nu e^{-i\omega_\lambda t}\right ] b_j^{\alpha\nu},
\end{eqnarray}
where the eigenvectors denoted by $\lambda$ appear in pairs for which the complex conjugate has a negative eigenvalue $-\omega_\lambda$; the eigenvectors/eigenvalues are ordered so the first $2N$ have positive eigenvalues and the latter $2N$ have negative eigenvalues.
Substituting this final expression into the equations of motion yields the following quadratic eigenvalue problem:
\begin{equation}
\label{eq:quadratic eigenvalue}
 \Big[{\bf M}\omega_{\lambda}^{2}+ i\omega_{\lambda}{\bf T}-{\bf K}\Big]{\bf Q}^\lambda=0,~\lambda=1,2,\ldots ,2N
\end{equation}
in which $\bf{M}$ is the diagonal mass matrix with ${\bf M}_{i\alpha,j\beta}=m_j\delta_{ij}\delta_{\alpha\beta}$, ${\bf T}$
is a real antisymmetric matrix defined by $T_{i\alpha,j\beta}=-eB_{eff}(j,\omega)\delta_{ij}\epsilon_{\alpha\beta}$ with $\epsilon_{xy}=-\epsilon_{yx}=1$ (and all other choices vanish), ${\bf K}$ is the (real symmetric) stiffness matrix in the
$xy$ plane given by
\begin{equation}
{\bf K}=\left (
\begin{array}{l l}
{\bf K}^{xx} & {\bf K}^{xy}\\
{\bf K}^{yx}& {\bf K}^{yy}
\end{array}
\right ),
\end{equation}
and the column eigenvector ${\bf Q}^\lambda$ satisfies $Q_{i\alpha}^\lambda=\sum_{\nu}\alpha_\lambda^\nu b_i^{\alpha\nu}$. In all cases, we are using a double index to represent the $2N\times 2N$ matrices, with $i$ running over indices $1,\ldots,N$, first with $\alpha=x$, and then with $\alpha=y$. We also assume that we are considering only the eigenvectors corresponding to nonnegative eigenvalues $\omega_\lambda\ge 0$.

We begin our analysis by defining a new set of matrices $\bar{\bf T}=m_{\rm ave}{\bf M}^{-1/2}{\bf T}{\bf M}^{-1/2}$ and $\bar{\bf K}=m_{\rm ave}{\bf M}^{-1/2}{\bf K}{\bf M}^{-1/2}$, and a new vector $\bar b^{\alpha\nu}={\bf M}^{1/2}b^{\alpha \nu}/\sqrt{m_{\rm ave}}$. The quadratic eigenvalue problem then becomes
\begin{equation}
 \Big[m_{\rm ave}\omega_{\lambda}^{2}\mathbb{I}+i\omega_{\lambda}{\bar {\bf T}}-{\bar {\bf K}}\Big]{\bar {\bf Q}^\lambda}=0,~\lambda=1,\ldots 2N,
\end{equation}
where ${\bar Q}_{i\alpha}^\lambda=\sum_{\nu}\alpha_\lambda^\nu b_i^{\alpha\nu}\sqrt{m_i/m_{\rm ave}}$ is the $\lambda$th eigenvector and all eigenvalues $\omega_\lambda$ remain nonnegative. Next, we define the real orthonormal eigenvectors $\bar b^{\alpha\nu}$ to satisfy
\begin{equation}
\sum_{j\beta} \bar {\bf K}_{i\alpha,j\beta}\bar b_j^{\beta\nu}=m_{\rm ave}(\omega_0^\nu)^2\bar b_i^{\alpha\nu},~\nu=1,2,\ldots 2N,
\end{equation}
where $\omega_0^\nu\ge 0$ is the normal mode frequency for motion in a vanishing effective magnetic field.  Then, we form the matrix $\bar{ \bf B}$ from the column vectors $\bar b^\nu$ and we evaluate the quadratic eigenvalue problem in the basis of the eigenvectors of $\bar b^\nu$ by multiplying on the left by $\bar{ \bf B}^{-1}=\bar{\bf B}^T$ and inserting $\mathbb{I}=\bar{\bf B}\bar {\bf B}^{-1}$ before the vector $\bar {\bf Q}$. This gives 
\begin{equation}
 \Big[m_{\rm ave}\omega_{\lambda}^{2}\mathbb{I}+i\omega_{\lambda}{\bar {\bar {\bf T}}}-{\bar {\bar {\bf K}}}\Big]{\bar {\bar {\bf Q}}}^\lambda=0,~\lambda=1,2,\ldots 2N,
\label{eq: quad_plane}
\end{equation}
with ${\bar {\bar {\bf T}}}={\bar {\bf B}}^T{\bar {\bf T}}{\bar {\bf B}}$, 
${\bar {\bar {\bf K}}}_{\nu\nu'}=m_{\rm ave}(\omega_0^{\nu })^2\delta_{\nu\nu'}$, and
${\bar {\bar {\bf Q}}}={\bar {\bf B}}^T{\bar {\bf Q}}$. Since these matrices are in the basis of the eigenvectors of ${\bar {\bf K}}$, they use the eigenvector index $\nu$ for the components of the stiffness matrix, which is diagonal, of course. In addition, using the orthonormality of the $\bar b$ eigenvectors, one immediately sees that ${\bar{\bar {\bf Q}}}_\nu^\lambda=\alpha^\nu_\lambda$.

The eigenvalues of the quadratic eigenvalue problem are all real, and by taking the complex conjugate of Eq.~(\ref{eq: quad_plane}), it is easy to see that if $\omega_\lambda$ is the eigenvalue corresponding to the eigenvector $\alpha_\lambda$ then $-\omega_\lambda$ is the eigenvalue corresponding to $\alpha_\lambda^*$. Since the solution of a nontrivial quadratic eigenvalue problem is not obvious, we map it onto a conventional eigenvalue problem, by doubling the size of the matrices.  We do so in a symmetrized fashion, so that the resulting matrix is Hermitian (but complex), and the solution of the eigenvalue problem follows conventional linear algebra techniques. The associated linear eigenvalue problem is
\begin{eqnarray}
&~&\left (
\begin{array}{c c}
-i{\bar {\bar {\bf T}}}/m_{\rm ave} & ({\bar {\bar {\bf K}}}/m_{\rm ave})^{1/2}\\
({\bar {\bar {\bf K}}}/m_{\rm ave})^{1/2} & 0
\end{array}
\right )\\
&\times&\left ( 
\begin{array} {l}
\omega_\lambda\alpha_\lambda\\
({\bar {\bar {\bf K}}}/m_{\rm ave})^{1/2}\alpha_\lambda
\end{array}
\right )
=\omega_\lambda \left ( 
\begin{array} {l}
\omega_\lambda\alpha_\lambda\\
({\bar {\bar {\bf K}}}/m_{\rm ave})^{1/2}\alpha_\lambda
\end{array}
\right ),\nonumber
\end{eqnarray}
where $([{\bar {\bar {\bf K}}}/m_{\rm ave}]^{1/2}\alpha_\lambda)_\nu=\omega_0^\nu\alpha_\lambda^\nu$, since the
stiffness matrix is diagonal in this basis (the repeated $\nu$ does {\it not} imply a summation over $\nu$). It is straightforward to show that the solution of this linear
eigenvalue problem solves the original quadratic eigenvalue problem.  We organize the solutions so that the first $2N$ eigenvalues are positive and the second $2N$ eigenvalues are negative. Since the negative eigenvectors of the quadratic eigenvalue problem are related to the positive eigenvectors via complex
conjugation, we automatically know the negative eigenvectors once the positives are known. Note that working in the $b$ basis produces a diagonal form for the spring-constant matrix, which allows us to easily take the matrix square root.

The eigenvectors of the linear eigenvalue problem (of dimension $4N$) satisfy the conventional orthogonality and completeness relations of Hermitian eigenvalue problems, but we choose them not to be normalized, so that we can find more standard relations between the $\alpha_\lambda$ eigenvectors.  Following the approach of Refs.~\cite{georgetown,Baiko}, one can derive the following relations based on orthogonality and completeness:
\begin{eqnarray}
&~&\sum_{\nu=1}^{2N}\left (\omega_\lambda\omega_{\lambda'}+\omega_0^{\nu 2}\right )\alpha_\lambda^{\nu *}\alpha_{\lambda'}^\nu=\frac{\omega_\lambda}{\hbar m_{\rm ave}}\delta_{\nu\nu'}\\
&~&\sum_{\lambda:\omega_\lambda>0}\omega_\lambda(\alpha_\lambda^{\nu *}\alpha_\lambda^{\nu'}+\alpha_\lambda^\nu\alpha_\lambda^{\nu'*})=\hbar m_{\rm ave}\delta_{\nu\nu'}\\
&~&\sum_{\lambda:\omega_\lambda>0}(\alpha_\lambda^{\nu *}\alpha_\lambda^{\nu'}-\alpha_\lambda^\nu\alpha_\lambda^{\nu'*})=0\\
&~&\sum_{\lambda:\omega_\lambda>0}\frac{1}{\omega_\lambda}(\alpha_\lambda^{\nu *}\alpha_\lambda^{\nu'}+\alpha_\lambda^\nu\alpha_\lambda^{\nu'*})=\frac{1}{\hbar m_{\rm ave}\omega_0^{\nu 2}}\delta_{\nu\nu'}.
\end{eqnarray}
In the last three of these expressions, the second terms in the parenthesis arise from the negative eigenvalue solutions of the quadratic eigenvalue problem.

\begin{figure*}[htbp!]
  \centering
    \includegraphics[width=0.9\textwidth,clip=on]{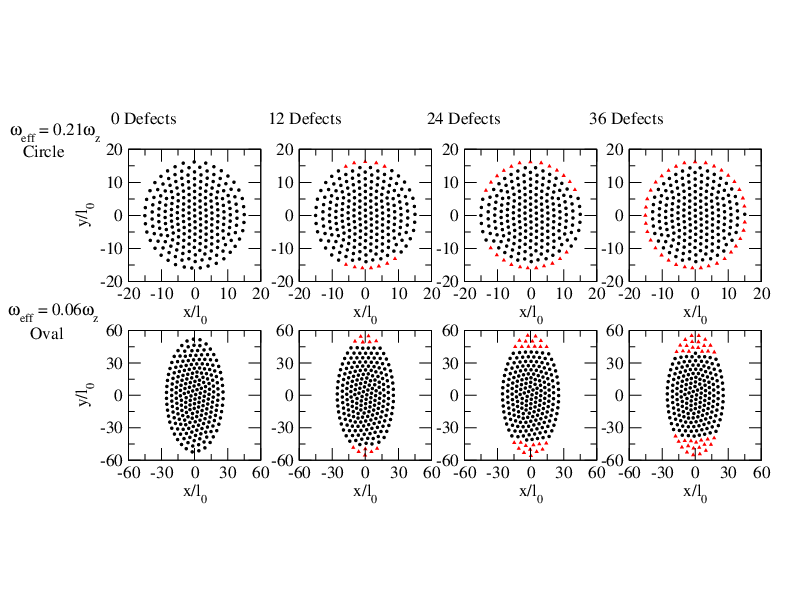}
\vskip -1.0in
\caption{(Color online.) Equilibrium structures for $N = 217$ ions and a weak wall ($\omega_W=0.04\omega_z$)
with different rotating frequencies. The top panels are the faster rotating circular case and the bottom panels the slower rotating oval case. The panels show $N_d=0$, 12, 24, and 36. Black circles denote Be$^+$ ions and red triangles denote BeH$^+$ molecules. Note the difference in the spatial scales for the upper and lower panels. }
\label{EqPos}
\end{figure*}

The quantization procedure for the planar modes is also nonstandard~\cite{Baiko,georgetown}. The complications arise because the generalized momentum operators do not commute with each other due to the magnetic field.  The procedure to properly quantize the Hamiltonian is still fairly standard, and we present only a sketch of the full derivation here.  We begin with the definition of the generalized coordinate
that we will be using. We let the displacement be represented by
\begin{eqnarray}
\delta R_j^\alpha(t)&=&\sum_{\nu=1}^{2N}\zeta_\nu(t)b_j^{\alpha\nu}\\
&=&\sum_{\nu=1}^{2N}\zeta_\nu(t)\bar b_j^{\alpha\nu}\sqrt{\frac{m_{\rm ave}}{m_j}},
\end{eqnarray}
where $\zeta_\nu$ is the generalized coordinate for $1\le\nu\le 2N$. As discussed above, the eigenvectors $b$ or $\bar b$ are both real, so the displacement is  real. The mechanical momentum satisfies $\Pi^\nu=m_{ave}\dot\zeta_\nu(t)$.  The corresponding generalized momentum is
\begin{equation}
P^\nu=\Pi^\nu-\frac{1}{2}\sum_{\nu'}\zeta_{\nu'}{\bar {\bar T}}_{\nu'\nu},
\end{equation}
where ${\bar {\bar T}}$ is the antisymmetric matrix that represents the effective magnetic field in the basis of the $\bar b$ eigenvectors, as described above. The classical Hamiltonian now takes a simple form in terms of the mechanical momentum and the generalized coordinate:
\begin{equation}
H_{ph}^P=\frac{1}{2m_{\rm ave}}\sum_{\nu}(\Pi^\nu)^2+\frac{1}{2}m_{\rm ave}\sum_{\nu}(\omega_0^\nu)^2(\zeta_\nu)^2.
\end{equation}
The problem with quantization is that we quantize with respect to the generalized momentum, not the mechanical momentum, so the mechanical momentum commutation relation becomes nonzero between different spatial directions. In other words, we take conventional commutation relations, where the $\hat \zeta$ operators commute with themselves, as do the $\hat P$ operators, and we take $[{\hat \zeta}_\nu,{\hat P}^{\nu'}]_-=i\hbar\delta_{\nu\nu'}$.  Then $[{\hat \Pi}^\nu,{\hat \Pi}^{\nu'}]_-=-i\hbar {\bar {\bar T}}_{\nu\nu'}$.  To complete the quantization, we construct the raising and lowering operators in terms of nonstandard combinations of the coordinate and mechanical momentum operators
\begin{eqnarray}
\hat a_\lambda^\dagger&=&\sum_{\nu=1}^{2N}\left (\alpha_\lambda^\nu \Pi^\nu+im_{\rm ave}\frac{(\omega_0^\nu)^2}{\omega_\lambda}\alpha_\lambda^\nu\zeta_\nu \right ) ,\\
\hat a_\lambda&=&\sum_{\nu=1}^{2N}\left (\alpha_\lambda^{\nu *} \Pi^\nu-im_{\rm ave}\frac{(\omega_0^\nu)^2}{\omega_\lambda}\alpha_\lambda^{\nu *}\zeta_\nu \right ) .
\end{eqnarray}
Using the relations satisfied by the $\alpha_\lambda$ eigenvectors, one can immediately verify that
$[\hat a_\lambda,\hat a_{\lambda '}^\dagger]_-=\delta_{\lambda\lambda'}$ (and the commutators of two $\hat a$'s or two $\hat a^\dagger$'s vanish). The quantized Hamiltonian then takes the standard form
\begin{equation}
\hat H^{P}_{ph}=\sum_{\lambda=1}^{2N}\hbar \omega_\lambda \left ( \hat a^\dagger_\lambda\hat a_\lambda +\frac{1}{2}\right ),
\end{equation}
which can be directly verified by substitution, or by checking that $[\hat H^P_{ph},\hat a^\dagger_\lambda]_-=\hbar\omega_\lambda\hat a_\lambda^\dagger$. Finally, the expression for the coordinate in terms of the raising and lowering operators becomes
\begin{equation}
{\hat \zeta}_\nu=-i\hbar\sum_{\lambda:\omega_\lambda > 0}\left ( \alpha_\lambda^{\nu *}\hat a_\lambda^\dagger-\alpha_\lambda^\nu \hat a_\lambda \right ).
\end{equation}
This relation is needed as the starting point to determine the effective spin-spin interactions, as it is needed to be substituted into the optical dipole force Hamiltonian which couples the spins to the positions of the ions~\cite{georgetown,spin-spin-interactions}.

Since there are a number of different ways that one might invoke an effective spin-spin coupling that employs driving the planar phonon modes, but none have been implemented experimentally yet, we do not examine effective spin-spin interactions in the planar directions any further in this work.  Of course, it would be a straightforward procedure to go from the quantized phonons to the spin-spin interactions when a spin-dependent force is applied, and this can be done in the future if it is needed for the case with defects in the ion crystal.

\section{Numerical results}
We present several numerical examples to illustrate the effect of defects on the equilibrium structure, 
eigenvalues and eigenvectors of the normal modes, and the spin-spin interaction $J_{ij}$ for 
the axial modes detuning above the center-of-mass mode. As in previous work, 
we use the axial center-of-mass phonon frequency (for the pure case) $\omega_z = 2\pi \times 795$ kHz (in units of rad/s) as the angular frequency unit. We define $\omega_W = \sqrt{2e|V_W|/m_{Be^+}}$ and only consider
the weak wall case, $\omega_W = 0.04\omega_z$. We use the Beryllium ion mass of $m_{Be^+}$ = 9.012182u
and BeH$^+$ ion mass of $m_{BeH^+}$ = 10.0201220u where u is the atomic mass unit. The magnitude of 
the magnetic field is $B_{z}=4.5$ Tesla corresponding to a cyclotron frequency 
$\omega_{c}(Be^+)=eB_{z}/m_{Be^+} = 9.645 \omega_z$ for the Beryllium ions and 
$\omega_{c}(BeH^+)=eB_{z}/m_{BeH^+} = 8.675 \omega_z$ for Beryllium hydride ions. We define the effective trapping frequency in the plane
as $\omega_{eff} = \sqrt{\omega_c \omega - \omega^2 - eV_0/m_{Be^+}}$ and only consider the effective frequencies
$\omega_h = 0.21\omega_z$ (rotation frequency $\omega=2\pi\times 45.3$~KHz for $\omega_z=2\pi\times 795$~KHz) and $\omega_l = 0.06\omega_z$ (rotation frequency $\omega=2\pi\times 41.3$~KHz for $\omega_z=2\pi\times 795$~KHz). We work with an ion crystal that has a total of $N=217$ particles, and allow it to have up to $N_d=$36 defects.

\subsection{Equilibrium configurations}
In Fig.~\ref{EqPos}, we plot the equilibrium positions of the ions when the system has a number of defects added for the two different rotation rates for the rotating wall potential.  One is a faster rotation rate, which tends to have closer-packed circular shaped crystals and the other is a slower rotation,
having a much larger sized crystal, with a marked oval asymmetry. Note how, in the circular case, the defects occupy the outer ring of the structure, while in the oval case, they are pushed primarily to the tips of the oval. Note how the spatial scale is three times larger in the bottom row (oval case). In general, while the equilibrium positions are certainly changed when defects are added, the change is not too large, and affects primarily the defect sites.

\subsection{Normal modes}
There are three branches of phonon modes.  The planar modes are described by the low-energy 
magnetron modes and the high energy cyclotron modes. The magnetron mode eigenvalues are hardly changed by the presence of defects, so we do not replot their values here. The cyclotron modes
have very little dispersion, and they are centered around the different cyclotron frequencies, hence $N_d$ of the eigenvalues are near $\omega_c(BeH^+)$, and $N-N_d$ eigenvalues are near
$\omega_c(Be^+)$. We show details of this behavior below, but first we show the eigenvalues for the axial mode frequencies, which can have large dispersion, and have a more complicated dependence on the number of defects.

\begin{figure}[h!]
  \centering
    \includegraphics[width=.48\textwidth,clip=on]{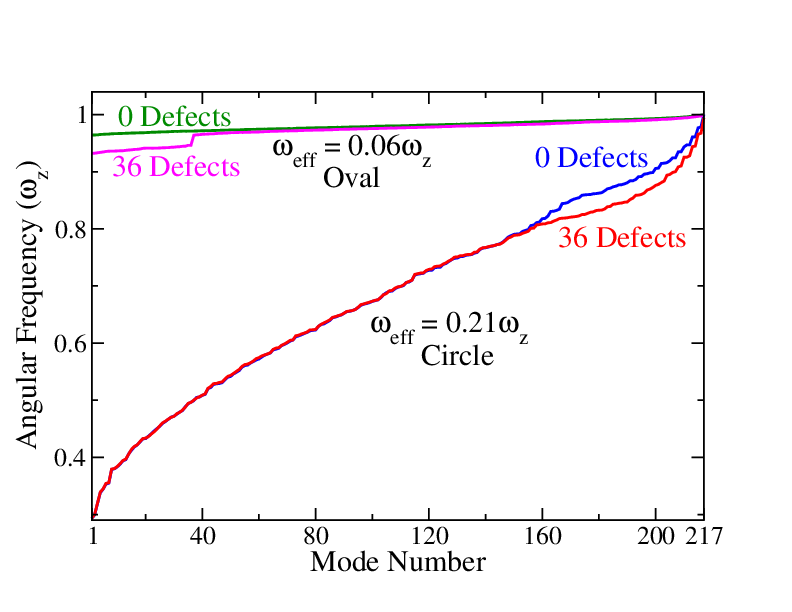}
\caption{(Color online.) Eigenfrequencies of the axial modes with defects for the fast and slow rotating cases. In general, there are about $N_d$ modes that are changed due to the defects, and these tend to be at the higher frequencies for the faster rotating case with a more circular profile and closer to the lower frequencies for the slower rotating oval case. The modes are plotted versus mode number, sorted according to increasing phonon frequencies.}
\label{Axial}
\end{figure}

The axial eigenvalues are plotted in Fig.~\ref{Axial} for the pure case and the case with $N_d=36$ defects. All modes clearly show an overall suppression of the phonon frequencies, while approximately $N_d$ of the modes are being suppressed by a larger amount (on the order of 3\%), which might be expected due to the mass difference. What is perhaps more surprising is that the $N_d$ modes that are suppressed more than the average ones occur at different regions of the spectrum when the shape of the equilibrium crystal changes.  In the circular case, they lie near the top of the spectrum and in the oval case near the bottom. The most probable explanation is that the isolated clustering of the defects in the oval case lends that system to have more localized modes that primarily involve the motion of defects only or of Be$^+$ ions only, which then shifts the lower energy frequencies due to the defects since the center-of-mass mode is the highest frequency mode for the axial phonons.  This effect is much reduced for the more circular crystals, where the defects are more evenly positioned around the perimeter of the crystal and the crystal is more homogeneous and closely packed, so that the largest change is for the collective modes that involve the motion of all ions and lie near the top of the spectrum.

\begin{figure}[htbp!]
  \centering
    \includegraphics[width=.48\textwidth]{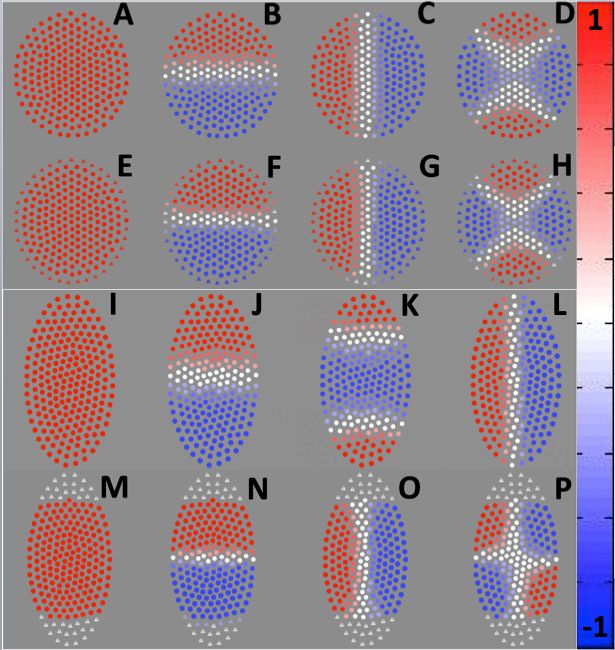}
\caption{(Color online.) False color image of the first four eigenvectors for the pure (top) and $N_d=36$ defect case (bottom) for the fast rotating (a-h) and slow rotating (i-p) cases.  Note how the eigenvectors are quite similar for the cases without (upper, a-d; i-l) and with (lower, e-h; m-p) defects. Be$^+$ ions are plotted with circles, BeH$^+$ defects with triangles. A careful inspection shows that the center-of-mass mode (far left) is no longer uniform for the defect case. Note that the defects can also change the ordering of the eigenvectors when sorted according to their eigenvalues, as seen by comparing panels (j-l) to (n-p). The color scale for the false color image is on the right.}
\label{AxialEigs}
\end{figure}

\begin{figure}[htbp!]
  \centering
    \includegraphics[width=.48\textwidth]{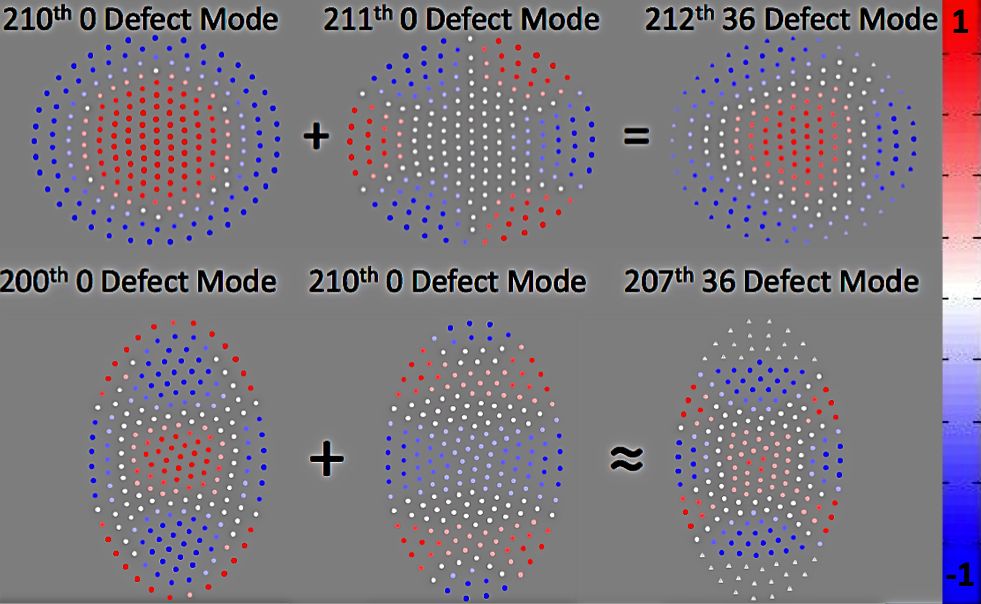}
\caption{(Color online.) False color image showing how one of the defect modes is related to simple linear combinations of different pure modes with similar eigenvalues.  The top panel is for the fast rotating circular case where a sum of the two pure case modes is essentially equal to the defect mode and the bottom for the slower rotating oval case where the sum of the two modes is approximately equal to the defect mode. }
\label{fig: lin_comb}
\end{figure}

A false color schematic of the four highest eigenvalues is plotted in Fig.~\ref{AxialEigs}. One can see that the eigenvectors do not have significant change for these four cases, although the center-of-mass mode is clearly no longer uniform. Cases that have more complex behavior do appear in the eigenvectors. In Fig.~\ref{fig: lin_comb}, we can see some cases of this behavior. In the top panel, we show how one of the modes for the defect case for fast rotation, is almost exactly related to a linear combination of two nearby modes in the pure system, while for the defect case for slow rotation, in the lower panel, we can see that one of the defect modes is (roughly) related to a linear combination of two nearby pure modes. 

\begin{figure}[htbp!]
  \centering
    \includegraphics[width=.48\textwidth]{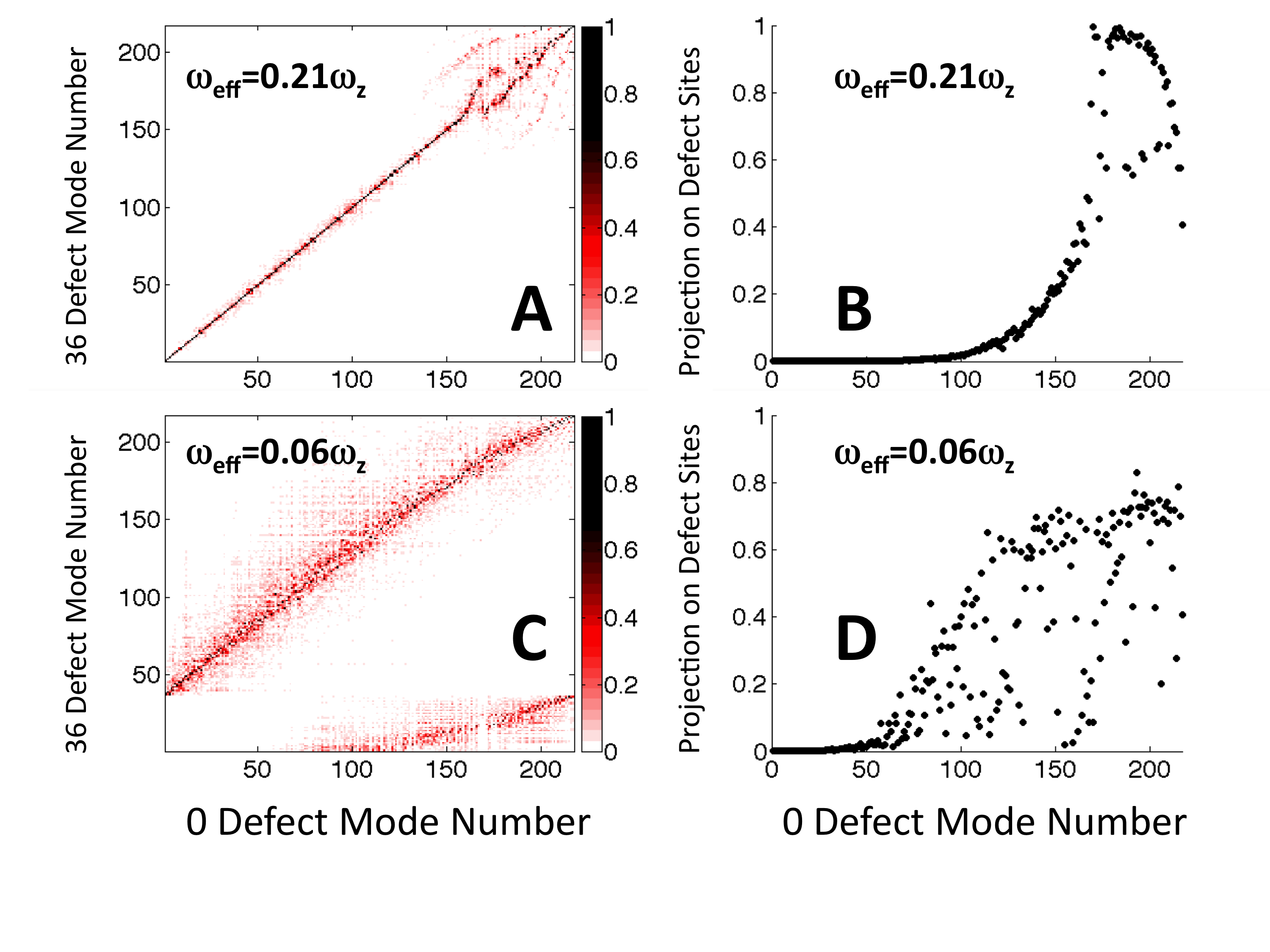}
\caption{(Color online.) False color image of the dot product of the pure-case eigenmodes with the corresponding defect-case eigenmodes ($\bar b$) with 36 defects (left) and norm of the pure-case eigenmode restricted to the defect sites (right).  Two cases are shown: (i) fast rotation (circular) case on the top and (ii) slow rotation (oval case) on the bottom.  }
\label{fig: mode}
\end{figure}

The fact that different phonon modes in the pure case are approximately combined into new modes for the defect case can be studied more quantitatively by determining the overlap of the defect eigenvectors onto the  eigenvectors for the pure case.  This is done computing the normalized ($\bar b$) eigenvectors for the relative displacements of the pure system, taking the dot product with the defect case eigenvectors and taking the absolute value.  We use the $\bar b$ eigenmodes because they form two orthonormal sets of vectors, so the expression is
\begin{equation}
{\rm Projection~on~\nu th~mode}=\left | \sum_i \bar b_i^{z\nu}(N_d=36)\bar b_i^{z\nu}(N_d=0)\right |
\end{equation}
We plot these results as a function of the mode numbers (sorted according to increasing phonon frequency)  in Fig.~\ref{fig: mode} with a false color image.  The fact that the results for the circular case (top)  lie predominantly on the diagonal indicates that the character of the modes does not change so much as the defects are added, but the effect becomes more severe near the higher modes, and this is precisely where the phonon frequencies were changed the most in the system. For the oval case (bottom), we see two high weight tracks and significant more spread amongst the results. This indicates that the defects are having a much more significant effect on the eigenvectors, but the second track on the subdiagonal indicates that the main effect is a reordering of the numbers of the modes, arising primarily from the phonons being more localized on the defect free and defect regions separately. In the right panels, we project the motion of the eigenvectors for the pure case onto the positions of the defects. This is done by first determining the $N_d=36$ positions for the equilibrium configuration that are farthest from the origin.  We project the pure case eigenvector onto those defect sites, and compute the norm of the projected vector. This norm is called the projection of the pure eigenmode onto the defect sites and is plotted in the right panels. One can see that there are far more modes with projections onto the defect sites for the oval case than for the circular case, and this explains why the eigenmodes have more mixing of the pure case modes in the oval case. Modes with little mixing on the defect sites will not have their frequencies changed by too much. These tend to be the low mode number localized modes of the axial eigenvectors.

\begin{figure}[htbp!]
  \centering
    \includegraphics[width=.48\textwidth]{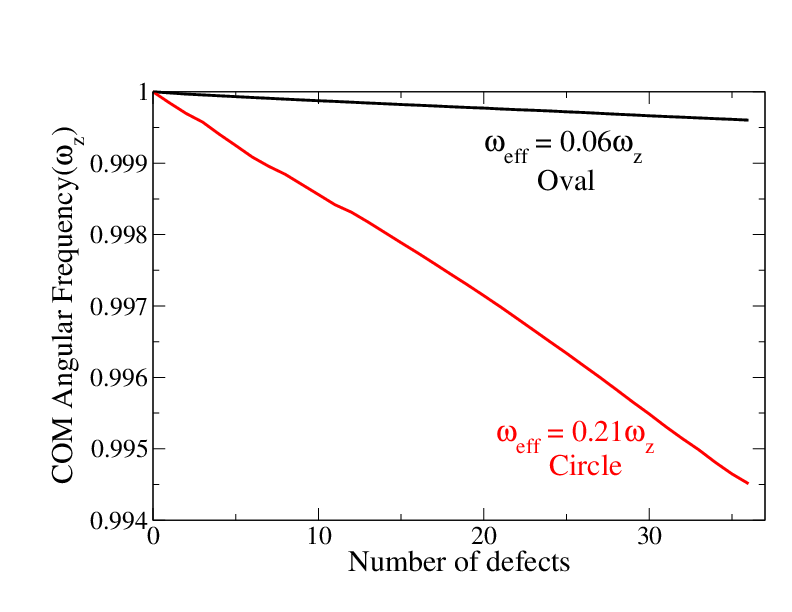}
\caption{(Color online.) Axial center-of-mass frequency as a function of the number of defects for the fast (black) and slow (red) cases. One can use these curves along with the measurement of the center-of-mass frequency to determine the number of defects in the system at any given moment of time (assuming the defects rapidly move to the boundary due to centrifugal effects).  }
\label{fig: com_defect}
\end{figure}

We next show how the center-of-mass mode changes with the number of defects for the two cases of fast and slow rotation in Fig.~\ref{fig: com_defect}. These curves are fairly linear, although structures in them can be clearly seen,  The slopes depend strongly on the shape of the crystal, being much smaller for the oval case. The relative shift of the frequency is small because the mass difference is only on the order of 10\% between the pure and defect sites.

\begin{figure}[htbp!]
  \centering
    \includegraphics[width=.45\textwidth]{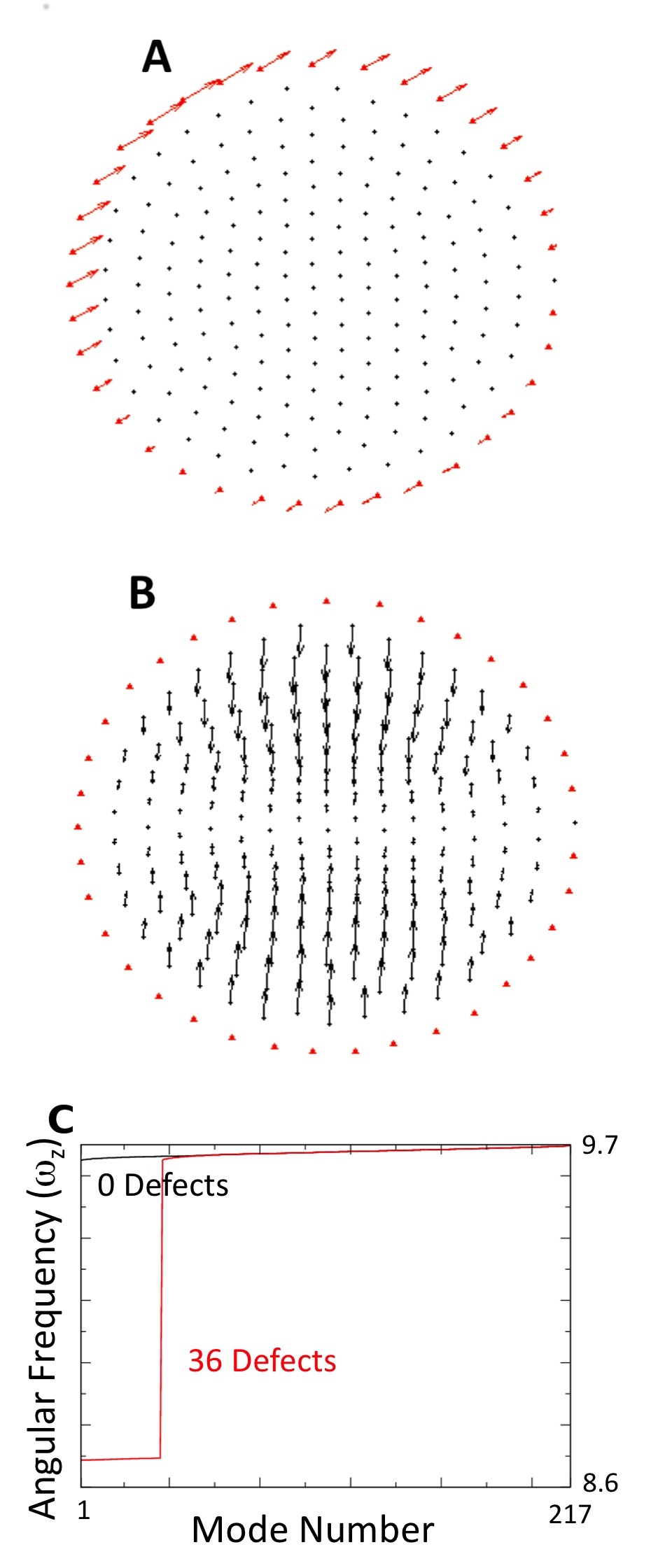}
\caption{(Color online.) Visualization of some example cyclotron modes (a) and (b) and of the mode spectrum (c) for $N_d=36$ and the fast rotating case; arrows indicate the instantaneous motion of the respective ionic site. One can see in panels (a) and (b) that the cyclotron modes separate into motion of the defects only (a) or of the Be$^+$ only (b).  The mode spectrum is shown in panel (c), where one can clearly see the separation of the motion according to the two different cyclotron frequencies for the defect case. }
\label{fig: cyclotron}
\end{figure}

We end this subsection by plotting some properties of the cyclotron planar modes in Fig.~\ref{fig: cyclotron}. Panels (a) and (b) show two selected modes of motion.  One can see that these modes are separated into motion of the defects only or motion of the Be$^+$ ions only, with very little coherent motion between the two species.  This occurs because the cyclotron motions of the two species are so different, much more so than the dispersion of the modes, and hence there is little collective motion between the two species. Example movies of these phonon modes are shown in the supplemental materials~\cite{supplemental}.

\begin{figure*}[htbp!]
  \centering
     \includegraphics[width=.45\textwidth]{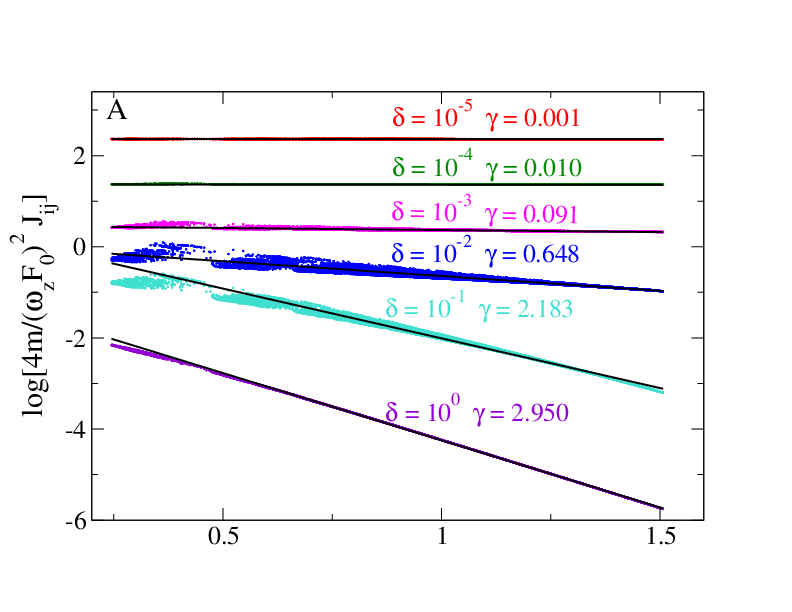}
 \includegraphics[width=.45\textwidth]{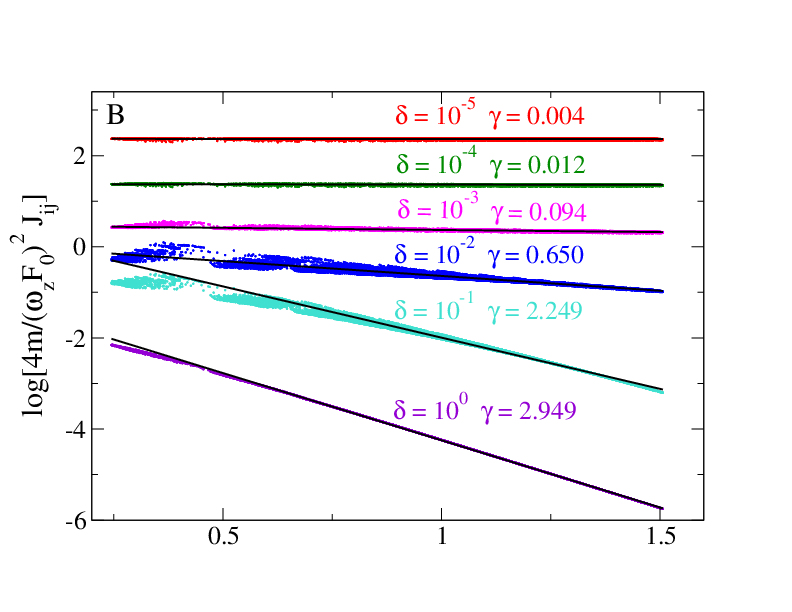}\\
\vskip -0.2in
 \includegraphics[width=.45\textwidth]{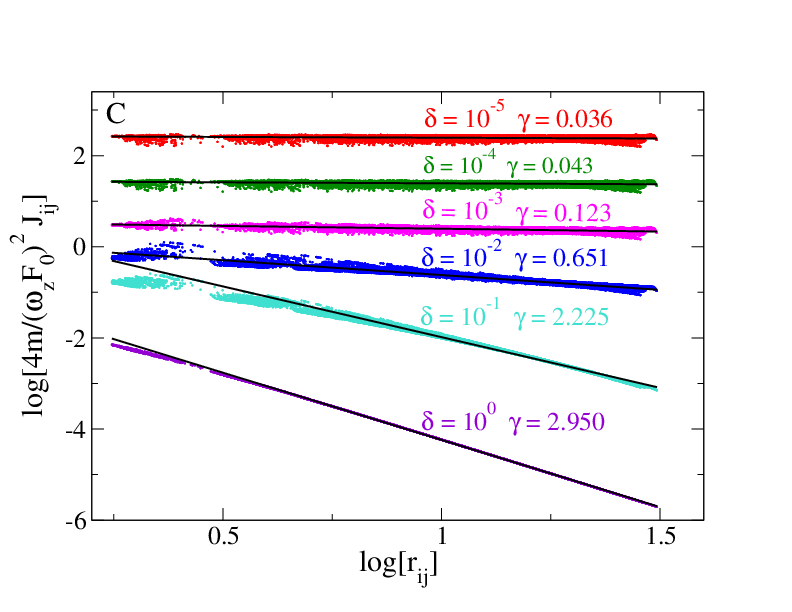}
 \includegraphics[width=.45\textwidth]{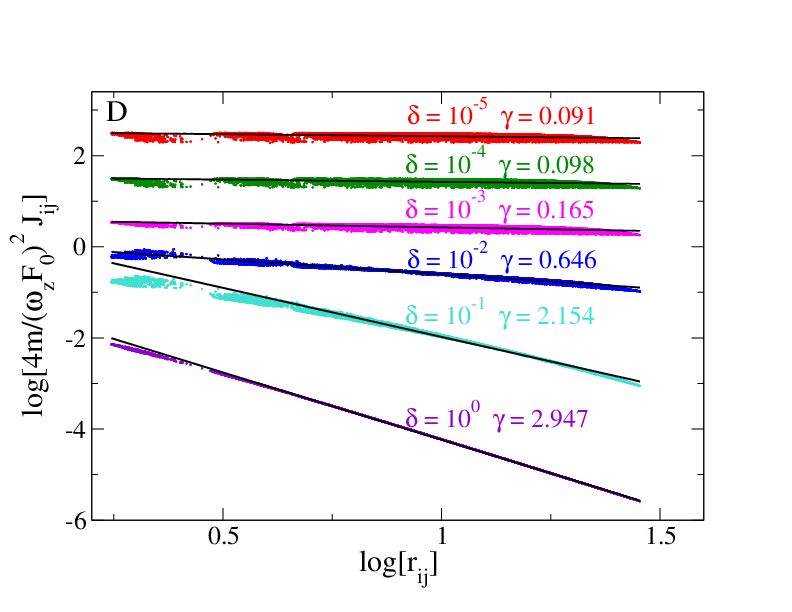}
\caption{(Color online.) Spin-spin couplings versus distance for different detunings (different colored lines) and different number of defects (different panels; a is 0 defects, b is 1 defect, c is 18 defects, and d is 36 defects) on a log-log plot. The straight line fits show power-law behavior $J_{ij}\propto r_{ij}^{-\gamma}$. The detunings are $\mu=\omega_{cm}+\delta\omega_z$, with $\delta=10^{-5}$, $10^{-4}$, $10^{-3}$, $10^{-2}$, $10^{-1}$ and 1. This case is for the faster rotating circular crystal.}
\label{fig: Jij}
\end{figure*}

\subsection{Ising spin-spin interaction for an axial drive}

Now that we have described how defects alter the phonon properties, we 
investigate how the spin-spin interactions compare to the pure case.   By coupling the phonons to the spins with a spin-dependent optical force, one can generate an Ising Hamiltonian ($\sigma_i^z$ is the Pauli spin matrix in the $z$-direction at site $i$)
\begin{equation}
\mathcal{H}=\sum_{ij} J_{ij}\sigma_i^z\sigma_j^z
\label{eq: ising}
\end{equation}
where we have neglected time-dependent terms, and the sum is only over pure sites where a Be$^+$ ion lies. This is because as defects are added to the system, those sites no longer have an addressable spin, and hence we only compute the spin-spin interactions between two sites that each contain a Be$^+$ ion.
Hence the spin-spin couplings become
\begin{equation}
J_{ij}=\frac{F_O^2}{4m_{ave}}\sum_{\nu=1}^N\frac{b_i^{z\nu}b_j^{z\nu}}{\mu^2-\omega_{z\nu}^2}
\label{eq: jfinal}
\end{equation}
for $i$ and $j$ Be$^+$ sites only.

To be specific, we examine detuning to the blue of the center-of-mass mode, correcting the detuning to lie to the blue of the center-of-mass mode with the defects. This way, all of the results will be compared on an equal footing.  Since the spin-spin couplings should behave like a power law in the distance between the ions, we fit the tails of the distributions to power laws.  The full range of the power law behavior ranges from 0 to 3, but because the center-of-mass mode is nonuniform when defects are present, the power law won't extend all the way down to zero.

\begin{figure*}[htbp!]
  \centering
     \includegraphics[width=.45\textwidth]{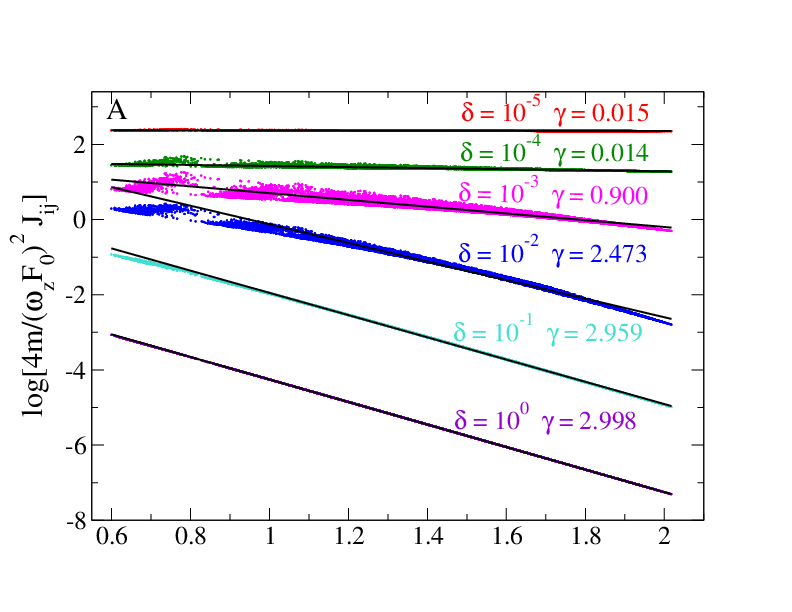}
 \includegraphics[width=.45\textwidth]{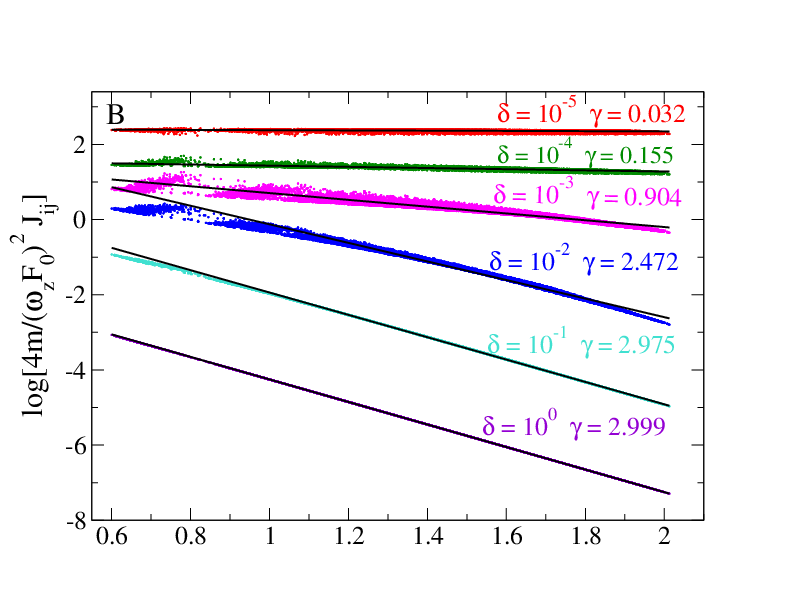}\\
\vskip -0.2in
 \includegraphics[width=.45\textwidth]{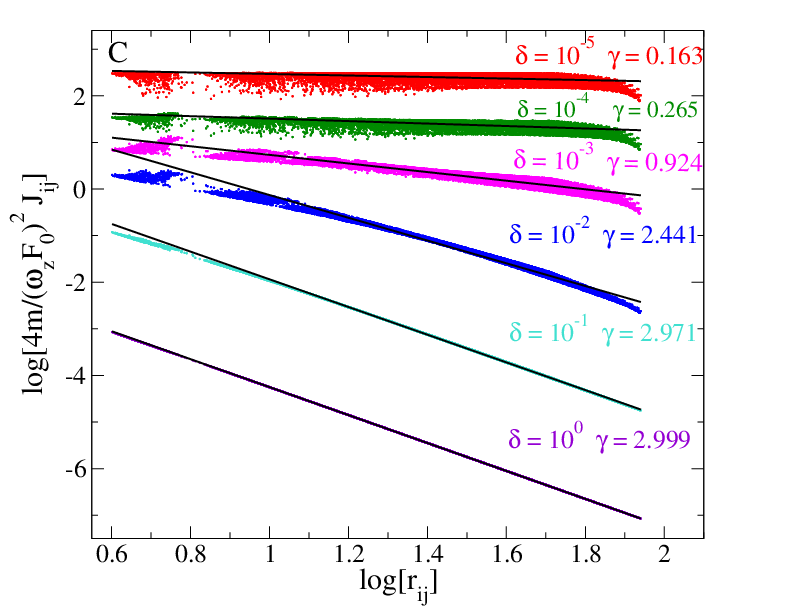}
 \includegraphics[width=.45\textwidth]{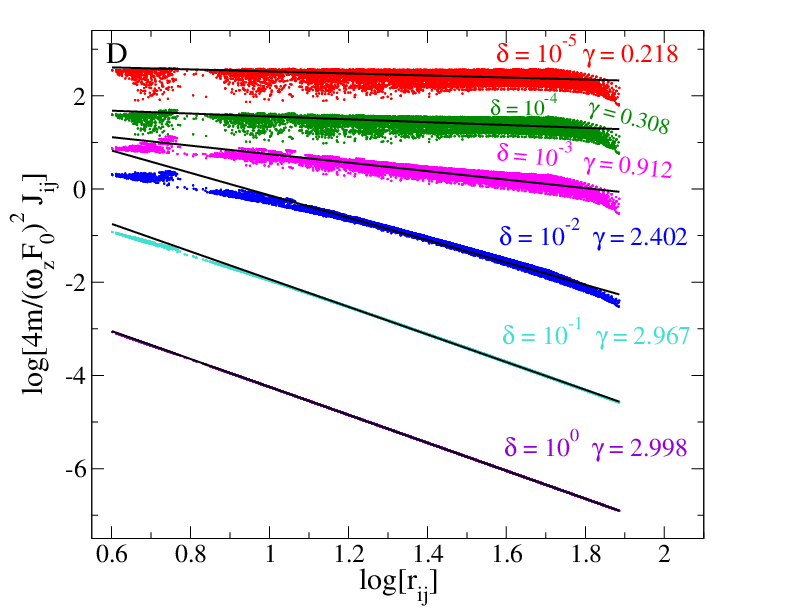}
\caption{(Color online.) Spin-spin couplings versus distance for different detunings (different colored lines) and different number of defects (different panels; a is for 0 defects, b is for 1 defect, c is for 18 defects, and d is for 36 defects) on a log-log plot. The straight line fits show power-law behavior $J_{ij}\propto r_{ij}^{-\gamma}$. The detunings are $\mu=\omega_{cm}+\delta\omega_z$, with $\delta=10^{-5}$, $10^{-4}$, $10^{-3}$, $10^{-2}$, $10^{-1}$ and 1. This case is for the slower rotating oval-shaped crystal.}
\label{fig: Jij_oval}
\end{figure*}

Results are shown in Fig.~\ref{fig: Jij} for the faster rotating circular crystal. One can immediately see that the main effect of the defects is to slightly increase the power law for an equivalent detuning, and for there to be a larger spread to the data as a function of distance when there are more defects. In particular, when there are enough defects, it is difficult to get to the point where the power law approaches zero. These results are quite important for experiment, as they show that the defects do not significantly alter the behavior of the spin-spin interactions, hence one does not need to maintain perfect purity when running a given experiment. The only issue that arises is one needs to ensure the detuning remains fixed relative to the center of mass mode, by keeping the number $\delta$ constant during the experiment [$\mu=\omega_{cm}+\delta\omega_z$], which does require carefully monitoring the center-of-mass mode frequency.  Not doing so, will give rise to a somewhat larger drift of the power laws as more defects are added to the crystal.

The situation is a bit different for the slower rotating oval crystal, depicted in Fig.~\ref{fig: Jij_oval}. Here, as the number of defects rises, we see two effects.  The first is an overall broadening of the width of the distribution of the couplings, especially for cases where the detuning is close to the center-of-mass mode, and we also see a long distance deviation of the power law behavior (also most pronounced for small detunings). These effects become worse as the number of defects rises, and must be due to the change in character of the phonon modes near the boundary region between the pure ions and the defects. Hence, the aspect ratio of the crystal plays a role in how well the spin-spin couplings are maintained as defects are added to the system, and as we expect, the problems are worse for close detuning to the center-of-mass mode.

\section{Conclusion and Discussion}

In this work we have developed a general theory and formalism to describe the phonon modes of a single-plane rotating Coulomb crystal in a Penning trap, and used those phonon modes to determine the relevant spin-spin couplings of the system. We find that, in general, the effects of the defects are rather mild even when they result in more than 15\% of the lattice sites in the crystal. The biggest effect is in modifying the character of some of the phonon modes, in a well-determined fashion. Even so, the effect on the effective spin Hamiltonian generated by a spin-dependent optical force applied to the crystal, is rather small, especially if the relative detuning $\delta$ is held constant during the experiment and the crystal is nearly circular. These results clearly indicate that Penning trap quantum simulators are robust to the introduction of a number of defects into the system, and hence one need not control the purity to high levels in order to use them as quantum emulators. But deviations are more important for certain shape crystals, implying that one should go through this type of analysis to ensure that the experiments remain robust against the effects of the defects.

\section{Acknowledgements}
We thank John Bollinger, Joe Britton,
and Brian C. Sawyer for valuable discussions. We also thank Adam Keith and C.-C. Joseph Wang, who worked on intial stages of this project and created Fig. 1. This work was initially supported under ARO
grant number W911NF13100019 with funds from the DARPA OLE Program and was completed under support from the NSF under grant number PHY-1314295. J. K. F. further
acknowledges support from the McDevitt bequest at Georgetown. B. Y. also acknowledges support from the Achievement Rewards for College Scientists Foundation.

\end{document}